\documentclass[%
reprint,
showpacs,preprintnumbers,
 amsmath,amssymb,aps,
]{revtex4-1}

\usepackage{graphicx}
\usepackage{dcolumn}
\usepackage{bm}
\usepackage{natbib}

\begin{document}

\def\H{\widehat}
\def\n{\widetilde{\nabla}}
\def\w{\widetilde{\delta}}
\def\K{\widehat{K}}
\def\R{\sqrt{K^r K_r}}
\def\D{\widetilde{\Delta}}
\def\N{\nonumber}

\preprint{GGG/FF-UV/2017}

\title{Covariant perturbations in the gonihedric string model}

\author{Efra\'\i n Rojas}
\email{efrojas@uv.mx}
\affiliation{%
Facultad de F\'\i sica, Universidad Veracruzana, Cto. Gonz\'alo 
 Aguirre Beltr\'an s/n, Xalapa, Veracruz 91000, M\'exico
}%





\begin{abstract}
We provide a covariant framework to study classically the stability of small 
perturbations on the so-called gonihedric string model by making precise use 
of variational techniques. The local action depends of the square root of the 
quadratic mean extrinsic curvature of the worldsheet swept out by the string,
and is reparametrization invariant. A general expression for the worldsheet 
perturbations, guided by Jacobi equations without any early gauge fixing, is 
obtained. This is manifested through a set of highly coupled nonlinear 
differential partial equations where the perturbations are described by scalar fields, 
$\Phi^i$, living in the worldsheet. This model contains, as a special limit, to 
the linear model in the mean extrinsic curvature. In such a case the Jacobi equations
specialize to a single wave-like equation for $\Phi$.
\end{abstract}

\maketitle


\section{\label{sec:intro}Introduction}

The so-called gonihedric string model is considered as a natural extension 
of the Feynman path integral over random walks, to an integral over random 
surfaces. The action functional is defined in a way that, when the worldsdsheet 
swept out by a closed string, degenerates into a worldline such an action must be 
reduced to an action of a point-like relativistic particle~\cite{Savvidy1992a,Savvidy1992b,Savvidy1993}. 
This field theory, at classical level, predicts the existence of tensionless 
strings possessing a massless spectrum of higher integer spin gauge 
fields~\cite{Savvidy2002,Savvidy2003,Savvidy2004} whereas, at quantum level,
fluctuations generate a nonzero string tension~\cite{Savvidy1992a,Savvidy1992b,Savvidy1993}. 
Moreover, when the theory is formulated on an Euclidean lattice it has a close 
relationship with a spin system which generalizes the Ising model with ferromagnetic, 
antiferromagnetic and quartic interactions~\cite{Savvidy1994,Savvidy1995,Savvidy2015}.

The effective action for the theory is proportional to the linear size of a 
random surface. Under this assumption, in building the model, a dimensional 
analysis as a guide unavoidably entails that the concept of extrinsic curvature 
associated to the surface must be included~\cite{Savvidy1992a,Savvidy1992b,Savvidy1993}. 
More specifically, on geometrical grounds, the model looks like an extrinsic volume 
through of the modulus of the mean extrinsic curvature, $K^i$, where $i$ stands 
for the number of normal vectors of the worldsheet immersed into a background spacetime. 
While the associated equations of motion (eom) resemble wave-like equations for a unit vector, 
$\K^i$, in the normal frame of the theory, this apparent advantage is deceptive since 
the $\K^i$ are derived objects constructed from the physical field variables, that is, 
the embedding functions $X^\mu$. In terms of these, the eom are of fourth-order, 
which are intractable unless a high degree of symmetry is consider on the surface 
be considered. Even if mathematically a solution exists for this type of surfaces, 
the possibility of it being a physical object in nature depends on whether it 
is stable or not. Regarding this point, an elegant description for examining the 
stability of a geometric object is provided by a manifestly covariant analysis.

This paper, motivated mainly from some recent reviews~\cite{Savvidy2015,Savvidy2017}, 
is devoted specifically to obtain a manifestly covariant expression to describe 
the classically perturbations in a relativistic surface described by the gonihedric 
string model. Unlike some perturbative analyzes~\cite{Savvidy2002,Savvidy2003,Fazio2003} 
for this model, our emphasis is on the worldsheet geometry by taking advantage of the 
inherent geometric structures defined on the worldsheet. The perturbation analysis we develop 
is about classical solutions with the advantage of bringing to the foreground how the 
covariance under rotations of the normals to the worldsheet enters into the game.
In my opinion, this issue was overlooked in several contributions and does not appear
to have been addressed before. Additionally, our analysis is performed without impose 
any early gauge fixing. 

Following the guideline for brane theories deformations, it is known that for an extended 
object of arbitrary dimension, $p$, in building local actions using invariants characterizing 
the geometry of the associated $(p+1)$-di\-men\-sio\-nal worldvolume, at least three ingredients 
are mandatory: invariance under reparametrizations of the worldvolume, 
diffeomorphisms invariance of the background spacetime, and, when $p+1 < N$, invariance 
under rotations of the normal vectors adapted to the worldvolume~\cite{Arreaga2000,Carter1996}. 
This last fact is necessary to identify that in the geometric extrinsic structure 
description, we have the presence of a connection that guarantee the covariance under 
worldvolume normal rotations. Needless to say, we must take into account these facts 
if we try to understand deeply the geometrical underpinnings of the gonihedric relativistic 
model. A classically equivalent action, close in spirit to the one developed by G. Savvidy
and collaborators, to describe tensionless strings was performed in~\cite{Mourad2004,Mourad2006}.

Our aim is twofold. First, we highligth the geometric content of the model. In particular, 
we emphasize the covariance of the description not only with respect to worldsheet 
diffeomorphisms but also with respect to local rotations of the normals to the worldsheet 
since we have a codimension of the worldsheet greater than one. Second, we obtain a consistent 
covariant derivation  of the linearized equations of motion about classical solutions and reach 
the Jacobi equations describing the perturbations of the worldsheet governed by the gonihedric 
string action. These are highly coupled partial differential equations for a multiplet of 
scalars fields, $\Phi^i$, with support on the worldsheet.

The paper is organized as follows. In Sect. \ref{gonih} we briefly review the 
gonihedric string model and emphasize the role that the extrinsic geometry plays 
in its description. After established the notation and conventions, in Sect. \ref{first} 
we perform the first variation of the action and identify the equations of motion.
In Sect. \ref{second} we achieve the second variation of the action and thereby we 
obtain the linearized equations of motion which are nothing but the conditions for 
stability of this type of surfaces. 
We conclude in Sec.~\ref{conclusions} with some comments of the work.

\section{The gonihedric string model}
\label{gonih}

The gonihedric string action is defined by the 
functional~\cite{Savvidy1992a,Savvidy1992b,Savvidy1993,Savvidy2002,Savvidy2003,Savvidy2004} 
\begin{equation}
S[X^\mu] = \alpha \int_m d^2x\,\sqrt{-g}\,\sqrt{K^{i}K_i},
\label{action}
\end{equation}
where $X^\mu$ represents the field variables which correspond to the 
embedding functions of the smooth worldsheet, denoted by $m$, swept 
out by a closed string when evolving in a Minkowski $N$-dimensional 
spacetime, denoted by $\mathcal{M}$, with metric $\eta_{\mu\nu} = \mbox{diag} 
(-1,1.1,\ldots, 1)$ where $\mu,\nu = 0,1,2,\ldots,N-1$. To specify the string 
trajectory we set $x^\mu = X^\mu (x^a)$, where $x^\mu$ are local coordinates in 
the spacetime, $x^a$ are coordinates on the worldsheet ($a,b = 0,1$), and $\alpha$ 
is a constant with appropriate dimensions. Further, $g=\textrm{det} (g_{ab})$ and $K^i$ stands for the determinant 
of the induced metric on $m$ and the trace of the $i$-th extrinsic curvature 
$(i,j= 1,2,\ldots,N-2)$, respectively, (see below for details).

The two vectors $e^\mu{}_a := \partial_a X^\mu$ form a basis of tangent vectors 
to $m$. These allow to define an induced metric on $m$ as $g_{ab} := e_a \cdot e_b$. 
Hereafter, a central dot will denote contraction with the Minkowski 
metric. In addition, $\nabla_a$ will denote the (torsionless) covariant derivative 
compatible with $g_{ab}$. Similarly, the $i$-th normal vector to $m$, $n^{\mu\,i}$, 
is defined by the relations
\begin{equation}
e_a \cdot n^i = 0, 
\qquad \qquad \qquad 
n^i \cdot n^j = \delta^{ij},
\end{equation}
where $\delta^{ij}$ is the Kronecker delta. It is worth to note that these expressions 
determine $n^{\mu\,i}$ only up to a $O(N-2)$ rotation (and a sign) and it transforms as a 
vector under normal frame rotations~\cite{Cheng1973,Dajczer1990}. Whereas the tangential
indices are lowered and raised with $g_{ab}$ and $g^{ab}$, respectively, the normal
indices are lowered and raised with $\delta_{ij}$ and $\delta^{ij}$, respectively.

The gradients of the orthonormal basis entail the definition of the extrinsic curvature
tensor, $K_{ab}^i = - n^i \cdot \nabla_a \nabla_b X = K_{ba}^i$. Apart from the extrinsic 
curvature, the extrinsic geometry of a surface, when the codimension is higher than one, 
is complemented with the extrinsic twist potential, $\omega_a^{ij}$, defined by 
\begin{equation}
\omega_a^{ij} := \nabla_a n^i \cdot n^j = - \omega_a^{ji}.
\label{twist}
\end{equation}
In this sense, the so-called Gauss-Weingarten equations describing an embedded
timelike worldsheet in a spacetime are given by
$\nabla_a e^\mu{}_b = - K_{ab}^i n^\mu{}_i$ and $\nabla_a n^{\mu\,i} = K_{a}{}^{b\,i} e^\mu{}_b
+ \omega_a^{ij} n^\mu{}_j$. In the case of a hypersurface embedding, $i=1$, the extrinsic twist vanishes 
identically. Under a rotation $n^i \rightarrow O^i{}_j n^j$, this potential 
transforms as a connection so that, this quantity is considered as the gauge 
field associated with the normal frame rotation group~\cite{Guven1993}. Therefore,
to implement normal frame covariance in a manifest way, we need the existence 
of a new covariant derivative $\n_a$, defined on fields transforming as tensors
under normal frame rotations~\cite{Cheng1973,Dajczer1990,Guven1993}, 
$\n_a \Phi^i{}_j := \nabla_a \Phi^i{}_j - \omega_a^{ik}\Phi_{kj} - 
\omega_{a\,jk} \Phi^{ik}$. This fact signals the presence of a curvature
associated with the twist defined by $\Omega_{ab}{}^{ij} := \nabla_b \omega_a^{ij} 
- \nabla_a \omega_b^{ij} - \omega_b^{ik} \omega_{a\,k}{}^j + \omega_a^{ik} 
\omega_{b\,k}{}^j$.

Some remarks are in order. First, it must be stressed that the action~(\ref{action}) is proportional 
to the lenght of the surface. Geometrically, this corresponds to the linear 
size of the surface, as it was for the path 
integral~\cite{Savvidy1992a,Savvidy1992b,Savvidy1993}. Second, in Ref.~\cite{Savvidy2003b} 
Savvidy considered the gonihedric model in two classically equivalent theories: The model
$A$, when the independent field variables are the embedding functions $X^\mu$, and
the model $B$ in which both $X^\mu$ and the induced metric $g_{ab}$ are the independent
variables. In this parlance, in our geometrical approach we will develop 
further the so-called model $A$.

\section{First variation and the equations of motion}
\label{first}

In order to pave the way to obtain the stability conditions for this type of surfaces
we opt to directly vary the main geometric quantities involved. The variation
of the action~(\ref{action}) is
\begin{equation}
\label{var1}
\delta S = \alpha \int_m d^2 x\, \left( \delta \sqrt{-g} \,\sqrt{K^iK_i}
+ \sqrt{-g}\, \delta \sqrt{K^i K_i} \right).
\end{equation}
Carrying out the variations under the integral sign we have first that $\delta \sqrt{-g}
= (\sqrt{-g}/2) g^{ab} \delta g_{ab}$. Regarding the second term we have 
$\delta \sqrt{K^i K_i} = \frac{K^i}{\sqrt{K^j K_j}} \w K_i$. Here, we have considered the 
deformation operator, $\w$, covariant under normal frame rotations, constructed in
analogy to the covariant derivative $\n_a$~\cite{Capovilla1995}. We requiere also that
$\w K^i = \w (g^{ab} K_{ab}^i) = \delta g^{ab} K_{ab}^i + g^{ab} \w K_{ab}^i = -
K^{ab\,i} \delta g_{ab} + g^{ab} \w K_{ab}^i$. These relations allow us to write
the variation~(\ref{var1}) in the form
\begin{eqnarray}
\label{var2}
\delta S &=& \alpha \int_m dA\, \left\lbrace \K_i \left[ \left( \frac{1}{2}
g^{ab} K^i - K^{ab\,i} \right) \delta g_{ab} \right. \right.
\nonumber
\\
&+& \left. \left.g^{ab} \w K_{ab}^i \right] \right\rbrace,
\end{eqnarray}
where the dependence of the variations of the first and second fundamental forms, 
$g_{ab}$ and $K_{ab}^i$, respectively, is explicitly manifested, and where we have 
also introduced the unit vector, $\K^i := \frac{K^i}{\sqrt{K^jK_j}}$. This vector 
may be thought of as being the coordinates of a $S^{(N-2)}$ unit sphere, $\delta_{ij} 
\K^i \K^j = 1$. Evidently, $\K^i$ transforms like a vector under normal frame rotations. 
Additionally, hereafter, we will use $dA:= d^2x\,\sqrt{-g}$, for short in the notation.

The infinitesimal changes of the field variables, $X^\mu(x^a) \to X^\mu (x^a) + \delta 
X^\mu (x^a)$, can be decomposed into tangential and normal deformations, that is, 
$\delta X^\mu = \Phi^a e^\mu{}_a + \Phi^i n^{\mu}{}_i$ where $\Phi^a$ and $\Phi^i$ denote 
both tangential and normal deformation fields, respectively. On the other hand, the fact 
that the gauge symmetry of the action~(\ref{action}) is the invariance under reparametrizations 
of the worldsheet, it determines that only the transverse worldsheet motion is 
physical so that the tangential deformations, $\delta_\parallel X^\mu = \Phi^a 
\partial_a X^\mu$, are usually ignored~\cite{Capovilla1995}. Therefore, we will only
consider $\delta_\perp X^\mu = \Phi^i n^\mu{}_i$, where the $\Phi^i$ are assumed to
be functions of $x^a$. Consequently, according to the geometric approach introduced in
Ref.~\cite{Capovilla1995}, the variations of the fundamental forms are given by the simple 
expressions 
\begin{eqnarray}
\label{id1}
\delta_\perp g_{ab} &=& 2K_{ab}^i \,\Phi_i,
\\
\label{id2}
\w_\perp K_{ab}^i &=& - \n_a \n_b \Phi^i + K_{ac}^i K^c{}_{b\,j}\Phi^j.
\end{eqnarray}
In terms of these, Eq.~(\ref{var2}) becomes
\begin{equation}
\label{var3}
\delta_\perp S = - \alpha \int_m dA\, \left[ \D \K^i - \left( g^{ab} K_j 
- K^{ab}_j \right) K_{ab}^i \K^j \right] \Phi_i,
\end{equation}
up to a total derivative where $\D = g^{ab} \n_a \n_b$ denotes the worldsheet 
d'Alembertian operator. Hence, the classical string trajectories are obtained
from the $N-2$ relations
\begin{equation}
\label{eom1}
\mathcal{E}^i = \D \K^i - J^{ab}_j K_{ab}^i \K^j = 0,
\end{equation}
where we have introduced the symmetric tensor
\begin{equation}
\label{Jab}
J^{ab}_i := g^{ab} K_i - K^{ab}_i.
\end{equation}
This is conserved in the sense that $\n_a J^{ab}_i = 0$ which is courtesy
of the Codazzi-Mainardi integrability condition for surfaces when the background
spacetime is Minkowski~\cite{Kobayashi69}. For future convenience, we also 
introduce the symmetric tensor in the normal frame
\begin{equation}
\label{Rij}
R_{ij} := J^{ab}_i K_{ab\,j} = R_{ji}.
\end{equation}
Clearly, $\mathrm{Tr}(R_{ij}) = \delta^{ij} R_{ij} = K^i K_i - K_{ab}^iK^{ab}_i
= \mathcal{R}$, the worldsheet Ricci scalar, which is nothing but the contracted
Gauss-Codazzi integrability condition in a flat spacetime background. In terms
of the tensor~(\ref{Rij}), the eom~(\ref{eom1}) can be written in the fashion
\begin{equation}
\label{eom2}
\mathcal{E}^i = \D\K^i - R^i{}_j \,\K^j.
\end{equation}
On pedagogical grounds, this set of equations can be seen as a set of wave-like
equations for the variables $\K^i$ but we must have in mind that the field 
variables are the embedding functions. Regarding this point, regrettably, the 
eom~(\ref{eom1}) are of fourth-order in the derivatives of $X^\mu$.

Hence, at this stage we have that under the deformation $X^\mu \to X^\mu + \delta X^\mu$,
the first variation of the action~(\ref{action}) reads
\begin{equation}
\delta_\perp S = -\alpha \int_m dA \, \left(\D\K_i - R_{ij} \,\K^j\right) \Phi^i, 
\label{res}
\end{equation}
where~(\ref{eom2}) is the result of the physical transverse motion provided by the 
breathing modes, $\Phi^i$, living in the worldsheet $m$.

Concerning the hypersurface embedding case, $i=1$, the action~(\ref{action}) 
specializes to a functional depending linearly of the mean extrinsic curvature
which has been discussed extensively in the relativistic context in the framework
of the Lovelock type branes~\cite{Rojas1,Rojas2} whereas in the Euclidean context
such functional has attracted lot of attention as being part of the geometrical
prescription to study biological lipid membranes~\cite{Capovilla2003,Capovilla2004,Rojas3,Rojas4}.
In such a case, $R_{ij}$ specializes to $\mathcal{R}$ so that, $\K^{(1)} = 1$
and the equations of motion~(\ref{eom2}) reduce to a single equation of second order
in the derivatives of the fields, $\mathcal{E}_{(1)} = \mathcal{R} = 0$. 
In other words, we recuperate the case of the action extremized by worldsheets
with vanishing Ricci scalar curvature.

\section{Second variation and the linearized equations of motion}
\label{second}

The way we follow to obtain the linearized equations of motion is to exploit
the relations given by~(\ref{id1}) and~(\ref{id2}). As discussed formally in 
Ref.~\cite{Capovilla2003}, the computation of the second 
variation of an action for surfaces simplifies enormously when the Euler-Lagrange 
equations are satisfied. There, has been proved that the calculation of the second 
variation is equivalent to a repeated application of the normal deformation operator, 
$\w_\perp$. A related approach about the analysis of the stability
for minimal surfaces was developed long time ago by H. A. Schwarz in \cite{Schwarz1890}. 
In this sense, from~(\ref{res}), let us consider the second variation of 
the action~(\ref{action})
\begin{equation}
\label{second1}
\delta^2 S_\perp = \int_m d^2 x\,\delta_\perp (\sqrt{-g} \,\mathcal{E}_i
\,\Phi^i),
\end{equation}
where, modulo the eom, it follows that the relevant equation to be exercised is
\begin{equation}
\label{var4}
\w_\perp \mathcal{E}_i = \w_\perp \left( \D \K_i - R_{ij}\,\K^j \right).
\end{equation}
Clearly, we note that three specific variations are involved,  $\w_\perp \K^i, \w_\perp R_{ij}$
and $\w_\perp (\D \K^i)$ 
so that we perform these in steps. Guided by~(\ref{id1}) and~(\ref{id2}) we get

\begin{eqnarray}
\label{id3}
\delta_\perp g^{ab} &=& - 2 K^{ab}_i\,\Phi^i,
\\
\label{id4}
\w_\perp K^{ab}_i &=& - \n^a \n^b \Phi_i - 3 K^{(a}{}_{c\,i} K^{b)c}_j\,\Phi^j,
\\
\label{id5}
\w_\perp K^i &=& - \D \Phi^i + (R^i{}_j - K^i K_j )\Phi^j,
\end{eqnarray}
where we have considered the fact that $K_{ab\,i} K^{ab}_j = - R_{ij} +
K_i K_j$. Hence, it is straightforward to compute the variation of the unit
vector $\K^i$
\begin{equation}
\w_\perp \K^i = - \frac{1}{\R} \Pi^{ij} \left( \D \Phi_j - R_{jl} \Phi^l \right),
\label{id6}
\end{equation}
where we have introduced the projection operator
\begin{equation}
\label{Pi}
\Pi^{ij} := \delta^{ij} - \K^i \K^j,
\end{equation}
satisfying 
\begin{equation}
\label{id7}
\Pi^i{}_j \K^j = 0 \qquad \mbox{and} \qquad \Pi^i{}_j \n_a \K^j =
\n_a \K^i.
\end{equation}
These results together with the unit vector fact $\K_i \n_a \K^i= 0$, provide
that $\{ \n_a K^i,\K^i \}$ is an orthonormal basis for a unit sphere $S^{(N-2)}$. 
Notice that for the case of a hypersurface, $i=1$, we have that $\Pi^{ij}$
vanishes. Similarly, a forthright computation leads to
\begin{eqnarray}
\w_\perp J^{ab}_i &=& - g^{ab} \D \Phi_i + \n^a \n^b \Phi_i + 3 K^{(a}{}_{c\,i}K^{b)c}_j\,
\Phi^j 
\N
\\
&+& 2 K_i J^{ab}_j\,\Phi^j + g^{ab} R_{ij} \,\Phi^j - 3 g^{ab} K_i K_j \,\Phi^j.
\label{id8}
\end{eqnarray}

In turn, 
by considering~(\ref{id2}) and~(\ref{id8}), the variation $\w_\perp R_{ij}$ after 
a straightforward computation is
\begin{eqnarray}
\w_\perp R_{ij} &=& - 2 J^{ab}_{(i} \n_{|a} \n_{b|} \Phi_{j)}
+ 2K_{(i} R_{j)l}\,\Phi^l - 2 K_i K_j K_l \,\Phi^l
\N
\\
&+& 2 K^a{}_{b\,i} K^b{}_{c\,j}
K^c{}_{a\,l}\,\Phi^l,
\label{id9}
\end{eqnarray}
or, collecting the last three terms in terms of $J^{ab}_i$ we are finally led to
\begin{equation}
\label{id10}
\w_\perp R_{ij} = - 2 J^{ab}_{(i} \n_{|a} \n_{b|} \Phi_{j)} - 2 K^a{}_{b(i} J^{bc}_{j)}K_{ca\,l}
\,\Phi^l.
\end{equation}
On consistency grounds, let us focus attention in the case of $i=1$. In
such a case $\w_\perp R_{ij} \to \delta_\perp \mathcal{R} = - 2\mathcal{R}_{ab} K^{ab} \,\Phi$,
up to a boundary term, as expected~\cite{Capovilla1995}, where $\mathcal{R}_{ab} = K^i K_{ab\,i}
- K_{ac}^i K^c{}_{b\,i}$ is the worldsheet Ricci tensor.

Regarding $\w_\perp (\D\K^i)$, we will consider the general expression
to compute the deformation of the d'Alembertian operator applied to an arbitrary
normal frame vector, $\Psi^i$,~\cite{Capovilla1995}
\begin{eqnarray}
\w_\perp ( \D\K^i ) &=& \D ( \w_\perp \K^i ) 
- 2 \n_a \left( K^{ab}_j\,\Phi^j\,\n_b \K^i \right) 
\N
\\
&+& \nabla^a \left( K_j  \Phi^j\right) \n_a \K^i
+ 2 K^{ab[i} ( \n_a \Phi^{j]} ) \n_b K_j  
\N
\\
&+& 2 \n_a [ K^{ab[i} ( \n_b \Phi^{j]} ) \K_j].
\label{id11}
\end{eqnarray}
Using the contracted Codazzi-Mainardi integrability condition 
we can put (\ref{id11}) as
\begin{eqnarray}
\w_\perp ( \D\K^i ) &=& \D ( \w_\perp \K^i ) 
- 2 K^{ab}_j \n_a\K^i \n_b \Phi^j 
\N
\\
&+& 2K^{ab\,i} \n_a \K_j \n_b \Phi^j - 2 K^{ab}_j \n_a \K^j \n_b \Phi^i 
\N
\\
&+& \n^a K^i \,\K_j\,\n_a \Phi^j - \n^a K_j \,\K^j \n_a \Phi^i  
\N
\\
&+& 2 K^{ab\,[i} \n_a \n_b \Phi^{j]}\,\K_j - 2 K^{ab}_j \n_a \n_b \K^i \,\Phi^j 
\N
\\
&+& \n^a \K^i\,K_j\,\Phi^j - \n_a \K^i \n^a K_j \,\Phi^j. 
\label{id12}
\end{eqnarray}

By collecting the variations~(\ref{id6}),~(\ref{id10}) and~(\ref{id12}) 
into~(\ref{second1}) we obtain
\begin{widetext}
\begin{eqnarray}
\delta^2 _\perp S &=& \int_m dA\, \left[ \w_\perp ( \D\K_i ) - \w_\perp R_{ij}
\,\K^j - R_{ij} \w_\perp \K^j \right] \Phi^i,
\N
\\
&=& \int_m dA \, \,\Phi^i \left[  
\D ( \w_\perp \K_i ) 
- 2 K^{ab}_j \n_a\K_i \n_b \Phi^j + 2K^{ab}_i \n_a \K_j \n_b \Phi^j 
- 2 K^{ab}_j \n_a \K^j \n_b \Phi_i  + \n^a K_i \,\K_j\,\n_a \Phi^j 
\right.
\N
\\
&-& \left. \n^a K_j \,\K^j \n_a \Phi_i + 2K^{ab}_{[i} \n_{|a} \n_{b|} \Phi_{j]} \,\K^j 
- 2 K^{ab}_j \n_a \n_b \K_i \,\Phi^j + \n^a \K_i\,K_j\,\Phi^j 
- \n_a \K_i \n^a K_j \,\Phi^j
\right.
\N
\\
&+& \left. 2J^{ab}_{(i} \n_{|a} \n_{b|} \Phi_{j)} \,\K^j
+ 2 K^a{}_{b\,(i} J^{bc}_{j)} K_{ca\,l} \K^j \,\Phi^l 
+ \frac{1}{\R} R_{il}\Pi^{lj} 
\left( \D \Phi_j - R_{jm} \Phi^m \right) \right].
\label{second2}
\end{eqnarray}
\end{widetext}
From now on, the factor $(-\alpha)$ will be omitted in the writing 
of the work, for short. To continue, we analyze the first term in~(\ref{second2})
\begin{equation}
\delta^2_\perp S_1 := \int_m dA\,\Phi^i \,\D ( \w_\perp \K_i )
= \int_m dA\, \D \Phi^i\,\w_\perp \K_i ,
\N
\end{equation}
where we have neglected divergence terms. Inserting now the variation~(\ref{id6}) 
into this expression and continuing integrating by parts as well as ignoring the 
divergence terms, we get
\begin{eqnarray}
\delta^2_\perp S_1 &=& \int_m dA\,\Phi^i \left[ - \D \left( \frac{1}{\R} \Pi_{ij} \D \Phi^j
\right) \right. 
\N
\\
&+& \left. \frac{1}{\R} R_{il} \Pi^{lj} \D \Phi_j \right],
\label{id13}
\end{eqnarray}
Now, by substituting~(\ref{id13}) into~(\ref{second2}) we can write the second 
variation of the action as
\begin{eqnarray}
\delta^2 _\perp S &=& \int_m dA\, \Phi^i \left[ - \D \left( \frac{1}{\R} \Pi_{ij} \D \Phi^j
\right)
\right. 
\N
\\
&+& \left. \frac{2}{\R} R_{il} \Pi^{lj} \D \Phi_j 
- 2 K^{ab}_j \n_a\K_i \n_b \Phi^j 
\right. 
\N
\\
&+& \left. 2K^{ab}_i \n_a \K_j \n_b \Phi^j 
-  2K^{ab}_j  \n_a \K^j \n_b \Phi_i 
\right. 
\N
\\
&+& \left. \n^a K_i \,\K_j\,\n_a \Phi^j - \n^a K_j \,\K^j \n_a \Phi_i 
\right.
\N
\\
&+& \left. 2 K^{ab}_{[i} \n_{|a} \n_{b|} \Phi_{j]} \,\K^j - 2 K^{ab}_j \n_a \n_b \K_i \,\Phi^j 
\right. 
\N
\\
&-& \left. \n_a \K_i \n^a K_j \,\Phi^j + \n^a \K_i\,K_j\,\Phi^j 
\right. 
\N
\\
&+& \left. 2 J^{ab}_{(i} \n_{|a} \n_{b|} \Phi_{j)} \,\K^j  
+ 2 K^a{}_{b\,(i} J^{bc}_{j)} K_{ca\,l} \K^j \,\Phi^l 
\right. 
\N
\\
&-& \left.  \frac{1}{\R} R_{il}\Pi^{lj}R_{jm} \Phi^m \right]. 
\label{id14}
\end{eqnarray}
In order to reduce this to a more familiar form, we will work some of the terms 
in the third and fourth lines of this expression
\begin{eqnarray}
 -2K^{ab}_j \n_a \n_b \K_i &-& \n_a \K_i\,\n^a K_j = 2 J^{ab}_j \n_a \n_b \K_i
\N
\\
&-& \n^a (K_j \n_a \K_i) - K_j R_{il}\K^l,
\N
\end{eqnarray}
and
\begin{eqnarray}
2 ( J^{ab}_{(i} \n_{|a} \n_{b|} \Phi_{j)} &+& 2 K^{ab}_{[i} \n_{|a} \n_{b|} 
\Phi_{j]} ) \K^j 
\N
\\
&=& 2\delta_{ij} J^{ab}_l \K^l \n_a \n_b \Phi^j 
- \R \Pi_{ij}\D \Phi^j
\N
\end{eqnarray}
where we have considered the equations of motion~(\ref{eom2}). Substituting these 
into the variation~(\ref{id14}) yields
\begin{eqnarray}
\delta^2 _\perp S &=& \int_m dA \,\Phi^i \left[ - \D \left( \frac{1}{\R} 
\Pi_{ij} \D \Phi^j \right) 
\right.
\N
\\
&+& \left. \frac{2}{\R} R_{il} \Pi^{lj} \D \Phi_j - \R \,\Pi_{ij}\D \Phi^j
\right. 
\N
\\
&+& \left. 2 \delta_{ij} J^{ab}_l \n_a \K^l \n_b \Phi^j 
+ 2\delta_{ij} J^{ab}_l \K^l \n_a \n_b \Phi^j  
\right. 
\N
\\
&-& \left. \delta_{ij} \K_l \n^a K^l \n_a \Phi^j + 4 K^{ab}_{[i} \n_{|a|} \K_{j]} 
\n_{b} \Phi^j 
\right.
\N
\\
&+& \left. \n^a K_i \,\K_j \n_a \Phi^j 
+ \n^a \K_i \,K_j \n_a \Phi^j 
\right.
\N
\\
&+& \left. \left( 2 J^{ab}_i \n_a \n_b \K_j 
- \n^a K_j  \n_a \K_i  - G_{ij} \right)\Phi^j
\right]
\label{second3}
\end{eqnarray}
where, once again, we have used the equations of motion~(\ref{eom2}) and introduced 
the following tensor in the normal frame
\begin{equation}
\label{Gij}
G_{ij} := \frac{1}{\R} R_{il} \Pi^{lm} R_{mj} - 2K^a{}_{b(i} J^{bc}_{l)} K_{ca\,j} \K^l.
\end{equation}
The remarkable thing about the tensor $G_{ij}$ is that it does not contain 
derivatives neither of the unitary vector $\K^i$ nor the breathing deformation 
field $\Phi^i$. 
Now, an integration
by parts in the second line of Eq.~(\ref{second3}) yields
\begin{eqnarray*}
\int_m dA \Phi^i 2\delta_{ij} J^{ab}_l \n_a \K^l \n_b \Phi^j
&=&
\\
&-& \int_m dA \Phi^i \delta_{ij} J^{ab}_l \n_a \n_b \K^l \Phi^j
\end{eqnarray*}
up to a boundary term. The expression~(\ref{second3}) is finally 
rearranged in the covariant form
\begin{equation}
\label{id15}
\delta^2_\perp S = \int_m dA \,\Phi^i \mathcal{L}_{ij}\Phi^j,
\end{equation}
where the local differential operators $\mathcal{L}_{ij}$ are given by
\begin{widetext}
\begin{eqnarray}
\label{Jacobi1}
\mathcal{L}_{ij} \Phi^j &=& - \D \left( \frac{1}{\R} \Pi_{ij} \D \Phi^j
\right) + \frac{2}{\R} R_{il} \Pi^{lj} \D \Phi_j - \R \,\Pi_{ij}\D \Phi^j
+ 2 \delta_{ij} J^{ab}_l \K^l
\n_a \n_b \Phi^j
\N
\\  
&-& \delta_{ij}  \K_l \n^a K^l \n_a \Phi^j  
+ 4 K^{ab}_{[i} \n_{|a|} \K_{j]} \n_{b} \Phi^j 
+ \n^a K_i \,\K_j \n_a \Phi^j + \n^a \K_i \,K_j \n_a \Phi^j 
\N
\\
&+& \left( 2 J^{ab}_i \n_a \n_b \K_j - \delta_{ij}\,J^{ab}_l \n_a \n_b \K^l
 - \n^a K_j \n_a \K_i  - G_{ij} \right) \Phi^j.
\end{eqnarray}
\end{widetext}
From another point of view, the second-order variation~(\ref{id15}) can 
be seen as a related deformation $X^\mu \to X^\mu (x^a) + \Phi^{'i} n^\mu{}_i$ 
applied to the first variation of the action, and subsequently imposing $\Phi^{'i} 
= \Phi^i$. The system~(\ref{Jacobi1}) consist of $N-2$ highly non-trivial, coupled 
partial differential equations for the $\Phi^i$. Unfortunately, we face with 
fourth-order differential equations for $\Phi^i$ where the analytical solutions 
are hard to obtain. These expressions comprise what are known as Jacobi equations 
for the case of the gonihedric string field theory. 

Some aspects of these Jacobi equations are in order. Geometrically, this set of 
relations describes the behaviour of this type of surfaces which are close or 
in the neighborhood  of a reference surface one. These equations are explicitly 
covariant under local normal frame rotations. On the other hand, on physical grounds, 
the solutions for this set of equations address the question of stability through the 
nature of the breathing modes, $\Phi^i$, of the worldsheet.  There is the hunch that some 
of these Jacobi equations are pure gauge or not contribute to the stability analysis 
since we originally have a second-order derivative theory which gives rise to some 
spurious geometric degrees of freedom. This fact will supported by a constraint 
Hamiltonian analysis for the model~\cite{Rojas5}. Anyhow, this idea deserves further 
attention. 
We would like to go one step further in the understanding of the stability for 
surfaces when this type of nontrivial rigidity is present, but at this stage we do 
not have strong geometric arguments to provide an answer due to the high degree of 
complexity of the equations~(\ref{Jacobi1}). In fact, we expect that an unstable behaviour 
predominates due to the close relationship of this model with a Nambu-Goto model, 
in an extended space~\cite{Savvidy2005}, which is known to have the characteristic 
of being unstable. This issue also deserves special care.


One immediate approximation is given by $\n_a K^i = 0$. In this case we 
have $R_{ij} \K^j = J^{ab}_i K_{ab\,j} \K^j = 0$. It is believed that this limit may 
have impact on the short wavelength fluctuations around classical trajectories given 
by $K_j K^j K^i = K_{ab}^i K^{ab}_j K^j$. For this case the Jacobi equations~(\ref{Jacobi1}) 
read
\begin{eqnarray}
\mathcal{L}_{ij} \Phi^j &=& - \frac{1}{\R} \Pi_{ij} \D \D \Phi^j + 
 \frac{2}{\R} R_{il} \Pi^l{}_j \D \Phi^j 
\N
\\
&-& \R \Pi_{ij} \D \Phi^j + 2 \delta_{ij} J^{ab}_l \K^l \n_a \n_b \Phi^j 
\N
\\
&-& G_{ij} \Phi^j.
\label{Jacobi3}
\end{eqnarray}

As before, consistency in the full set of results is mandatory. To show this note 
first that the tensor~(\ref{Gij}) can be put in the fashion
\begin{eqnarray}
G_{ij} &=& \frac{1}{\R} R_{il} \Pi^{lm} R_{mj} + \R\,R_{ij} - K_i R_{jl}\K^l
\N
\\
&-& 2J^{ab}_i K_{bc\,j}K^c{}_{a\,l} \K^l.
\label{Gij2}
\end{eqnarray}
Then, for a hypersurface embbeding we get $G_{11} = -2 J^{ab}K_{a}{}^c K_{bc}$, where $J^{ab} = 
g^{ab} K - K^{ab}$. For this particular case, from~(\ref{Jacobi1}) we have
\begin{equation}
\label{Jacobi2}
\mathcal{L}\Phi = 2 (J^{ab} \nabla_a \nabla_b \Phi - J^{ab}K_{a}{}^c K_{bc} \Phi ) = 0.
\end{equation}
We have thus encountered a second-order differential equation for $\Phi$.
This equation is clearly in accord with the results found in Ref.~\cite{Rojas2} 
for the case of an action functional depending linearly of the trace of the extrinsic 
curvature which corresponds to the so-called second Lovelock type brane invariant. 
In another fashion, Eq.~(\ref{Jacobi2}) reads
\begin{equation}
\label{Jacobi4}
J^{ab} \nabla_a \nabla_b \Phi - M^2\,\Phi = 0,
\end{equation}
which looks like a type wave equation with a mass-like term of the form $M^2 :=
J^{ab}K_{a}{}^c K_{bc} = \mathcal{R}_{ab} K^{ab}$, with $\mathcal{R}_{ab}$
being the worldsheet Ricci tensor.

\section{Concluding remarks}
\label{conclusions}

In this paper we have presented a covariant approach for the analysis of 
perturbations on surfaces governed by the so-called gonihedric string field 
theory. We have mainly focused on the normal deformations since these represent 
the only physically perturbations. According to the Savvidy sorting, we opt to 
develop the Model A for this theory with the idea of maintain the original 
field variables and of exploiting the natural geometric structures associated 
to the worldsheet. Though the square root Lagrangian seems complicated, 
we can handle it and exhibit in an elegant fashion by means of the unit
vector $\K^i$. Regarding this point, owing to inner geometrical nature of the 
action~(\ref{action}), the introduction of the unit vector exhibits the 
extreme elegance and simplicity of the eom~(\ref{eom1}) and helps to 
simplify, in some sense, the form of the stability conditions for 
the model. The generalization of the action~(\ref{action}) to describe 
$p$-dimensional extended objects is straightforward and the geometric analysis 
can be carried out following similar lines obtaining analogous conclusions. In 
this regard, there is a related approach to discuss the concept of rigidity in 
quantum gravity in relation with the so-called gonihedric principle by describing 
the propagation of compact orientable random branes with no spatial boundary~\cite{Savvidy1996,Savvidy1997a,Savvidy1997b,Rojas6}.

\begin{acknowledgments}
Enlightening remarks and discussions with Miguel Cruz are acknowledged. This 
work was partially supported by Sistema Nacional de Investigadores, M\'exico. 
Also, partial support from ProDeP-2017-M\'exico is acknowledged. 
\end{acknowledgments}






\begin{thebibliography}{99}

\bibitem{Savvidy1992a} 
R. V. Ambartzumanian, G. K. Savvidy, K. G. Savvidy and G. S. 
Sukiasian, \textit{Phys. Lett. B} \textbf{275} (1992) 99-102.

\bibitem{Savvidy1992b}
G. K. Savvidy and K. G. Savvidy, \textit{Mod. Phys. Lett. A} \textbf{8} (1992)
2963-2971.

\bibitem{Savvidy1993}
G. K. Savvidy and K. G. Savvidy, \textit{Int. J. Mod. Phys. A} \textbf{8} (1993) 
3993-4011.

\bibitem{Savvidy2002} 
G. K. Savvidy and R. Manvelyan, \textit{Phys. Lett. B} \textbf{533} (2002) 138-145.
 
\bibitem{Savvidy2003} 
A. R. Fazio and G. K. Savvidy, \textit{Mod. Phys. Lett. A} {\bf 18} (2003) 2817-2828. 

\bibitem{Savvidy2004} 
G. K. Savvidy, \textit{Int. J. Mod. Phys. A} \textbf{19} (2004) 3171-3194.


\bibitem{Savvidy1994}
G. K. Savvidy and F. J. Wegner, \textit{Nucl. Phys. B} \textbf{413} 
(1994) 605-613.

\bibitem{Savvidy1995}
G. K. Savvidy, K. G. Savvidy and F. J. Wegner, \textit{Nucl. Phys. B} \textbf{443} 
(1995) 565-580.

\bibitem{Savvidy2015}
G. Savvidy, \textit{Mod. Phys. Lett. B} \textbf{29} (2015) 1550203.

\bibitem{Savvidy2017}
G. Savvidy, Gravity with linear action and gravitational singularities,
\textit{arXiv:1705.01459} [hep-th].

\bibitem{Fazio2003}
A. R. Fazio, \textit{Acta Phys. Pol. B} \textbf{34} 4825-4834 (2003).

\bibitem{Arreaga2000} 
G. Arreaga, R. Capovilla and J. Guven, \textit{Annals of Phys.} \textbf{279} 
(2000) 126-158.

\bibitem{Carter1996}
B. Carter, Brane dynamics for treatment of cosmic strings and vortons, in 
\textit{``Recent Developments in Gravitation and Mathematics'', Proc. 2nd 
Mexican School on Gravitation and Mathematical Physics}, eds.  A. Garcia, 
C. Lammerzahl, A. Macias, T. Matos and D. Nu\~nez.
(Science Network Publishing, Konstanz, 1997); hep-th/9705172.

\bibitem{Mourad2004}
J. Mourad, \textit{``Continous spin and tensionless strings''}, arXiv:hep-th/0410009.

\bibitem{Mourad2006}
J. Mourad, \textit{AIP Conf. Proc.} \textbf{861} (2006) 436-443.

\bibitem{Cheng1973}
B. Y. Cheng, \textit{Geometry of Submanifolds} (Dekker, 1973).

\bibitem{Dajczer1990}
M. Dajczer, \textit{Submanifolds and Isometric Immersions} (Publish or Perish, 1990).

\bibitem{Guven1993}
J. Guven, \textit{Phys. Rev. D} \textbf{48}, 4606 (1993).

\bibitem{Savvidy2003b}
G. K. Savvidy, \textit{Phys. Lett. B} \textbf{552} (2003) 72-80.

\bibitem{Capovilla1995}
R. Capovilla and J. Guven, \textit{Phys. Rev. D} \textbf{51} (1995) 6736-6743.

\bibitem{Kobayashi69}
S. Kobayashi and K. Nomizu, \textit{Foundations of Differential Geometry: Volume 
II} (Interscience, New York, 1969).

\bibitem{Rojas1}
M. Cruz and E. Rojas, \textit{Class. Quant. Grav.} \textbf{30} (2013) 115012.

\bibitem{Rojas2}
N. Bagatella-Flores, C. Campuzano, M. Cruz and E. Rojas, \textit{Class. Quant. Grav.}
\textbf{33} (2016) 245012.

\bibitem{Svetina1989}
S. Svetina and B. \v{Z}ek\v{s}, \textit{Eur. Biophys. J.} \textbf{17} (1989) 101-111.

\bibitem{Capovilla2003}
R. Capovilla, J. Guven and J. A. Santiago, \textit{J. Phys. A: Math. Gen.} \textbf{38}
(2003) 6281-6295.

\bibitem{Capovilla2004}
R. Capovilla and J. Guven, \textit{J. Phys. A: Math. Gen.} \textbf{37} (2004) 5983-6001.

\bibitem{Rojas3}
R. Capovilla, J. Guven and E. Rojas, \textit{J. Phys. A: Math. Gen.} \textbf{38}
(2005) 8201-8210.

\bibitem{Rojas4}
R. Capovilla, J. Guven and E. Rojas, \textit{J. Phys. A: Math. Gen.} \textbf{38}
(2005) 8841-8860.

\bibitem{Rojas5}
E. Rojas, \textit{Ostrogradsky Hamiltonian analysis of the gonihedric string theory}.
In preparation.

\bibitem{Schwarz1890}
H. A. Schwarz, \textit{Gesammelte Mathematische Abhandlungen. Erster Band.}
(Springer-Verlag Berlin Heidelberg, 1890).

\bibitem{Savvidy2005}
G. Savvidy, \textit{Phys. Lett. B} \textbf{615} (2005) 285-290.

\bibitem{Savvidy1996}
G. K. Savvidy and K. G. Savvidy, \textit{Mod. Phys. Lett. A} \textbf{11} (1996) 1379-1396.

\bibitem{Savvidy1997a}
J. Ambjorn, G. K. Savvidy and K. G. Savvidy, \textit{Nucl. Phys. B} \textbf{486} (1997)
390-412.

\bibitem{Savvidy1997b}
G. K. Savvidy, \textit{Nucl. Phys. B Proc. Suppl.} \textbf{57} (1997) 104-114.

\bibitem{Rojas6}
C. Campuzano, R. Capovilla, A. Cervantes and E. Rojas, \textit{AIP Conf.Proc.} \textbf{1420} 
(2012) 42-46.


\end{thebibliography}

\end{document}